\documentclass[fleqn,usenatbib]{mnras}

\usepackage[T1]{fontenc}
\usepackage{soul}

\DeclareRobustCommand{\VAN}[3]{#2}
\let\VANthebibliography\thebibliography
\def\thebibliography{\DeclareRobustCommand{\VAN}[3]{##3}\VANthebibliography}

\usepackage{graphicx}	
\usepackage{amsmath}	
\usepackage{mwe}
\usepackage{subfig}
\usepackage{float}
\usepackage{enumitem}
\usepackage[none]{hyphenat}
\usepackage{adjustbox}

\usepackage[dvipsnames]{xcolor}

\newcommand{\kms}{km~s$^{-1}$}

\usepackage{newtxtext,newtxmath}

\title[2.5-MHD models of circumstellar discs]{2.5-MHD models of circumstellar discs around FS CMa post-mergers : I. Non-stationary accretion stage}

\author[Moranchel-Basurto et al.]{
A. Moranchel-Basurto,$^{1}$\thanks{E-mail: amoranchel087@gmail.com}
D. Korčáková,$^{1}$ and R. O. Chametla$^{1}$ \\
$^{1}$Charles University, Faculty of Mathematics and Physics, Astronomical Institute, V Hole$\check{s}$ovi$\check{c}$k\'ach 747/2, 180 00 Prague 8, Czech Republic\\
 }

\date{Accepted XXX. Received YYY; in original form ZZZ}

\pubyear{2023}

\begin{document}
\label{firstpage}
\pagerange{\pageref{firstpage}--\pageref{lastpage}}
\maketitle

\begin{abstract}
We investigate the dynamic evolution of gaseous region around FS~CMa post-mergers. Due to the slow rotation of a central B-type star, the dynamics is driven mainly by the magnetic field of the central star. Recent observations have allowed us to set a realistic initial conditions such as, the magnetic field value ($B_\star\approx6\times10^{3}G$), the mass of the central star ($M_\star=6M_\odot$), and the initial disc density $\rho_{d0}\in[10^{-13}\mathrm{g\,cm^{-3}},10^{-11}\mathrm{g \, cm^{-3}}] $. We use the PLUTO code to perform 2.5D-MHD simulations of thin and thick discs models. Especially relevant for the interpretation of the observed properties of FS~CMa post-mergers are the results for low-density discs, in which we find  formation of a jet emerging from inner edge of the disc, as well as the formation of the so called "hot plasmoid" in the corona region. Jets are probably detected as discrete absorption components in the resonance lines of FS~CMa stars. Moreover, the magnetic field configuration in the low-density plasma region, favors the appearance of magnetocentrifugal winds from the disc. The currents toward the star created by the magnetic field may explain accidentally observed material infall.
The disc structure is significantly changed due to the presence of the magnetic field. The magnetic field is also responsible for the formation of a hot corona as observed in several FS~CMa stars through the Raman lines. Our results are  valid for all magnetic stars surrounded by a low density plasma, i.e., some of stars showing the B[e] phenomenon.
\end{abstract}



\color{black}
\section{Introduction}

Recent studies on accretion discs around different magnetized stellar-type objects (e.g., classical T Tauri stars, white dwarfs, neutron stars, X-ray pulsars, magnetic cataclysmic variables) have focused on the effect of the stellar magnetic field on the inner rim structure during the "quiescent initial conditions", that means, in a stage where the accretion in the disc is stationary \citep{Kluzniak_Rappaport2007,Koldoba_etal2002,Romanova_etal2002,Long_etal2005,Bessolaz07,Zanni2009,ZF2013,Ireland_etal2021}. Most of the studies consider the case of rotating central  star  with a magnetic field at its surface. The diference between each of this models are mainly the initial conditions, for instance the strenght of magnetic field, the position of the magnetosphere, the radius or mass of the central star. However, similar results are found in these studies on the dynamics of the inner rim of the accretion disc, among which are: (1) The magnetospheric radius (where magnetic field lines are closed) must be
near the radius of corotation ($R_c$, where the disc material is stationary in the frame of the rotating star). In
larger radii, the field lines are deformed or inflated, and at the radius of corotation, the magnetosphere can be strongly modified by the disc. In other words, the radius of truncation is less than the corotation radius $R_c=(GM_\star/\Omega_\star^2)^{1/3}$, here $M_\star$ and $\Omega_\star$ are the mass and angular velocity of the star, respectively. (2) The presence of the funnel flux effect.  This effect occurs at the  disc-star interaction region where matter flows out of the plane of the disc and essentially falls freely along the stellar magnetic field lines. (3) Due to the relative speed of rotation between the central object and the accretion disc, the magnetic field is twisted. In fact, while the magnetic field is not radially distorted by the accumulation of matter due to the large radial magnetic diffusivity, the magnetic field is frozen in the disc in the azimuthal direction. Therefore, except at the corotation point, an azimuthal component $B_\phi$ of the magnetic field is created.

Recently, a new type of objects where the disc can be significantly 
influenced by the stellar magnetic field has been discovered.  The strong magnetic field ($\sim 6.2$~kG), kinematic properties, and
spectral properties of IRAS~17449+2320 \citep{Korcakova22} opened the possibility of the existence of
post-mergers
among FS~CMa stars, a sub-group of the B[e] stars.
A distinctive characteristic in
the spectra of the B[e] stars is the existence of forbidden emission lines and strong IR excess. 
These spectral properties point to the presence of very extended circumstellar region and dusty region. 
The B[e] phenomenon shows some of supergiants, Herbig Ae/Be stars, symbiotic stars, or compact planetary nebulae \citep{Lamers1998}. 
While the physical model of these four groups has been, at least at main points, explained, the nature and evolutionary status
of FS~CMa stars remains an open question. 

The properties of IRAS~17449+2320 suggest that there may be a group of intermediate-mass
mergers hidden among the FS~CMa stars. We use the term ``intermediate-mass mergers'' here to stress that mergers of lower-mass B-type stars are completely out of the focus of the community. The detailed simulations of the merger are done mostly for massive O-type
stars, because the product finally explodes as a supernova. Another type of merger in focus are Ap stars. 
Even if it seems that this gap is only a minor point, recent simulations of \cite{Nela_N-body} show that the majority of mergers are among the B-type stars. This may have important consequences for the enrichment of the ISM and stellar evolution, and rises the importance of the study of FS~CMa stars.

Motivated by what was described above, we decided to use
the results reported by \cite{Korcakova22} to explore this group of intermediaries massive mergers between FS~CMa stars through numerical simulations to investigate the dynamics of the inner rim and the disc corona. The values of the mass and the magnetic field of IRAS 17449+2320 are used as initial conditions. In a series of papers we analyze what would be the answers under this new scenario to what was discussed in points (1)-(3), and also in the case when the accretion on the star is not stationary. The latter is the central objective of this work, since the non stationary accretion can occur from different causes, for instance, in the early stage of evolution of young stellar objects or in a post-merger event which also applies here. The paper is organized as follows.  In the next section \ref{sec:observations}, some properties and characteristics that have been found observationally in FS~CMA stars are briefly described. In Section \ref{sec:model} we present the physical model, while the numerical implementations used in our 2.5D magnetohydrodynamical simulations are discussed in section  \ref{sec:numerical}. In Section \ref{sec:results} we present the results of our numerical models. A brief discussion is presented in Section \ref{sec:discussion} and finally concluding remarks can be found in Section \ref{sec:conclusions}.

\section{Observed properties of FS~CMa stars}
\label{sec:observations}

This section reviews some of the main observational features that have been found in FS~CMa objects. This group of objects 
 is defined based on the properties of the infrared excess 
\citep{Miroshnichenko_warm_dust_01, Miroshnichenko_FS_CMa}. However, this is not a sufficiently sharp criterion. 
The list of FS~CMa stars contains several binaries, e.g., Be/X-ray binary \citep[CI~Cam,][]{Barsukova06AR}. 
Other examples of binary systems are: 3-Pup, GG-Car, AS 386, MWC~300, MWC~349A.
The regular orbital periodicity was found in 3~Pup \citep{Miroshnichenko20_3Pup} and
GG~Car \citep{Porter21}. The amplitude of the radial velocities of AS~386 is very large pointing to 
a black hole as a secondary \citep{Khokhlov18}. MWC~300 is the nearest B[e] supergiant \citep{Appenzeller77}. It was found that the dust formation is due to the wind-wind (or wind-atmosphere) interaction in MWC~349A 
\citep{Tafoya04}. 
Several post-AGB stars are also in the list of FS~CMa stars; 
Hen3-938 \citep{Miroshnichenko99_Herbig}, 
CPD-48~5215 \citep{Gauba04}, and
maybe also MWC~939 \citep{Gauba03}.
The spectra of FS~CMa stars are very similar to those of Herbig Ae/Be stars. Therefore, they appear frequently
in lists of Herbig Ae/Be stars. However, a simple criterion works here; the star is or is not located in the star-forming region. 

In this study, we are interested in  another group hidden among FS~CMa stars, post-mergers. Currently, only one star was identified
as a post-merger \citep[IRAS 17449+2320,][]{Korcakova22}. However, the properties of several of them fit the merger
scenario. In the following, we summarise common properties of FS~CMa stars for which the post-AGB and pre-main sequence phase
is not likely and where the secondary has not been proven. However, only
a few members of this kind have been systematically monitored and have sufficient quality data. The following list is mainly based on the studies of FS~CMa itself, HD~50138, MWC~342, IRAS~17449+2320, and MWC~623. MWC~623 may seem disputable because the spectrum of this star
is composed of two components; hot B-type and cold K-type. Dealing with an ordinary main-sequence, it star would be strong evidence of the binary nature. However, FS~CMa stars are surrounded by a very extended and optically thick disc. The spectrum of the cold component may originate in the disk \citep{2028_Cyg_bisector, Zickgraf89}. Moreover, this star has extremely strong
\ion{Li}{i} resonance doublet 6~708~\AA, which supports the post-merger nature, because the overabundance of lithium 
has been measured in many of the red novae \citep{lithim_red_novae}. 

The common properties of stars listed above are:
\begin{itemize}[leftmargin=*]

  \item \textit{Gaseous highly inhomogeneous disc around the star.} Small moving absorption/emission humps are frequently observed in Balmer lines. This effect  studied well \cite{Pogodin97} in a series of spectra of HD~50138. He interpreted it as rotating structures. Since the data were obtained only during four days and the movements of structures have been clearly detectable, the rotating structures had to be close to the star. More observations with the   same conclusion have been taken also by \cite{Dachs92} and \cite{Terka16}. The presence of a gaseous disk follows also from the observations in resonance lines of \ion{Na}{i} D1, D2 and \ion{Ca}{ii} H and K lines. These lines show almost always the broad emission, indicating a low-density region above the disc. The disc is not purely Keplerian, but expands with small velocity, usually not exceeding 200 km/s.
 
 \item \textit{Geometrically thick disc.} Analysis of the interferometric data of FS CMa \citep{Hofmann22} show that the outer dust ring has an inclination angle around 40 deg. However, a~ strong absorption of iron-group elements is detected in UV spectra pointing to a~huge amount of gas also under such angles. 
 
 \item \textit{Expanding layers, that may decelerate} \citep{Kucerova13}.

 \item \textit{Clumps moving away from the star} \citep{Terka16}.

 \item \textit{Material infall} \citep{Kucerova13}.

 \item \textit{Rotating structures; clumps or arms.} Moving humps are observed frequently in every FS~CMa object for which the sufficiently high-resolution spectra are available.

 \item \textit{Iron curtain, i.e., the strong absorption of the iron group elements in the UV} \citep{Ivan-Pariz}. This is observed in stars seen from the equator or at the intermediate angles. The absorption may reduce the UV flux by about an order of magnitude compared to the classical Be stars. 

 \item \textit{UV excess} (detected in stars seen from the pole-on) \citep{Korcakova22}.  

 \item \textit{Slow rotation} (see appendix \ref{sec:appendix}). During the merger, the magnetic field is generated. As stronger the magnetic field is, as faster the star slows down \citep{Schneider19}.
 
 \item \textit{Most of them have large space velocities} (IRAS~17449+2320, \citealt{Korcakova22}, FS~CMa, HD~50138  \citealt{Nela_N-body}). The large space velocity may indicate the merger origin, because binaries that left the cluster, merge soon afterwards. The binary is kicked from the cluster due to the accidentally interaction with another star. In such a case, the eccentricity is almost always very large and the semi-major axis very small. Due to the tidal circularization, the semi-major axis is even shorter and components may merge. However, the large space velocity is not the necessary condition \citep{Nela_N-body}.

  \item \textit{Upper density and turbulent velocity limit of the outer parts.}
    Based on the Inglis-Teller formula (\citealt{Inglis_Teller, Nissen}), \cite{Kricek} found for FS~CMa
    the ion density $\sim 1\times 10^{11} \mathrm{cm}^{-3}$ and turbulent velocity $v_{\rm{turb}}<40$ km/s. 
    The limited values for IRAS~127449+2320 are $\sim 2\times 10^{12}\mathrm{cm}^{-3}$ and the turbulent velocity to $\sim$ 130~km/s \citep{Korcakova22}. 
\end{itemize}
  There are more common properties, e.g., the temperature and composition of the dust. However, 
    more details are beyond the scope of this paper, where
    we aim to describe the matter distribution around these stars and specify the influence of the magnetic field on the observed properties.

\section{Physical Model}
\label{sec:model} 
Our physical model is composed of a slightly sub-Keplerian disc rotating around a magnetized star (with a dipolar magnetic field configuration) as well as non-rotating 
corona initially in hydrostatic equilibrium.

Since we are interested in studying the physical effects that occur in the internal part between the rotating disc, the corona and the central star, we neglect the rotation of the central star, i.e., it acts as a point source of gravity and has a corona that possesses a magnetic dipole field at the center, which is perpendicular to the equatorial plane. This situation causes the system to be out of equilibrium in a short time. To support this physical model, in observational studies it has been found that the rotation velocities of FS~CMa type stars are low (see table \ref{vrottab} in the appendix). The physical model and numerical setup used in this study is based in some aspects in those given in \citep{Bessolaz07} and \citet{Zanni2009}, which we briefly recall for convenience.

\subsection{MHD equations}

The magnetohydrodynamic equations governing the gas dynamics are the conservation of mass
\begin{equation}
    \frac{\partial \rho}{\partial t}+\nabla \cdot (\rho \mathbf{v})=0,
\label{eq:mass}
\end{equation}
the conservation of momentum equation
\begin{equation}
    \frac{\partial \rho \mathbf{v}}{\partial t} + \nabla \cdot \left[\rho \mathbf{v} \mathbf{v} +\left(P+\frac{\textbf{B}\cdot \textbf{B}}{8\pi}\right)\mathbf{I}-\frac{\mathbf{B}\mathbf{B}}{4\pi}\right]=-\rho \nabla\Phi,
\label{eq:momentum}
\end{equation}
the energy equation
\begin{equation}
\begin{split}
   & \frac{\partial E}{\partial t} + \nabla \cdot \left\{ \mathbf{v} \left(E+P+ \frac{\mathbf{B}\cdot\mathbf{B}}{8\pi}\right) - \frac{1}{4\pi}[(\mathbf{v}\cdot \mathbf{B})\mathbf{B}+ \eta \textbf{J}\times \mathbf{B}]\right\} \\
   & = -\rho(\nabla\Phi) \cdot \mathbf{v},
\end{split}   
\label{eq:energy}
\end{equation}
and the induction equation,
\begin{equation}
    \frac{\partial \mathbf{B}} {\partial t}- \nabla\times (\mathbf{v} \times \mathbf{B} - \eta \textbf{J})=0.
    \label{eq.induction}
\end{equation}
where $P$ is the thermal gas pressure, $\rho$ the gas density, $\mathbf{v}$
its velocity, $\textbf{I}$ the unit tensor, $\textbf{B}$ the magnetic flux density vector, $\textbf{J}$ the electric current density vector, $\eta = 4\pi\nu_m$  the ohmic diffusion coefficient,
$\Phi = GM_*/R$ the gravitational potential and
$E$ represents the total energy density given by
\begin{equation}
    E=\frac{P}{\gamma-1}+\rho\frac{\mathbf{v}\cdot\mathbf{v}}{2}+\frac{\mathbf{B}\cdot\mathbf{B}}{8\pi}.
\end{equation}

In our model, the effects of viscosity and radiative cooling are neglected,  but we consider a magnetic resistivity $\nu_{m}$ given by \citep{Bessolaz07}:
\begin{equation}
    \nu_{m} = \alpha_{m} \Omega_K H^2 \exp\left(-\left(\frac{z}{H}\right)^4\right) ,
\end{equation}
which decreases with a disc scale height $H=C_s/\Omega_K$, with $C_S$ the sound speed, $\Omega_K$ the Keplerian rotation speed and $\alpha_m = 0.1$. We have taken the $z$-axis parallel to the rotation axis of the disc and assume that the magnetic dipole moment is aligned with the rotational axis. 

\subsection{Accretion disc}
\label{sec:accretion_disc}
The density $\rho_{d}$ and pressure $P_{d}$ of the gas in the accretion disc is set following the three-dimensional models of Keplerian accretion discs considering spherical coordinaties $(R,\theta,\phi)$ used in \citet{Zanni2009}:
\begin{equation}
    \rho_d (R,r) = \rho_{d0}\left[\frac{2}{5h^2}\left[\frac{R_0}{R}-\left(1-\frac{5h^2}{2}\right)\frac{R_0}{r}\right]\right]^{3/2}
    \label{eq:rho_}
\end{equation}

\begin{equation}
    P_d = h^2\rho_{d0}v_{K0}^2\left(\frac{\rho_d}{\rho_{d0}}\right)^{5/3}
\end{equation}
where $h=C_s/v_K$ is the aspect ratio, 
$\rho_{d0}$ is the initial disc density, and $v_{K0}= \sqrt{GM_*/R_0}$ the Keplerian velocity at $R=R_{0}$. In Eq. (\ref{eq:rho_}) $r=R\sin{\theta}$ is the cylindrical radius.

The velocity components of the accretion disc are calculated as in \cite{Bessolaz07}. The initial vertical component ($u_{\theta d}$) is zero, and since we do not consider any type of viscosity inside the disc, the radial component of the velocity is $u_{Rd}=0$. Lastly, the initial azimuthal component is set by the following expression:

\begin{equation}
    u_{\phi d} = \left(\sqrt{1-\frac{5h^2}{2}}\right)\sqrt{\frac{GM_\star}{r}},
\end{equation}
that is, we are considering a slightly sub-Keplerian disc.

\begin{figure}
\includegraphics[width=\columnwidth]{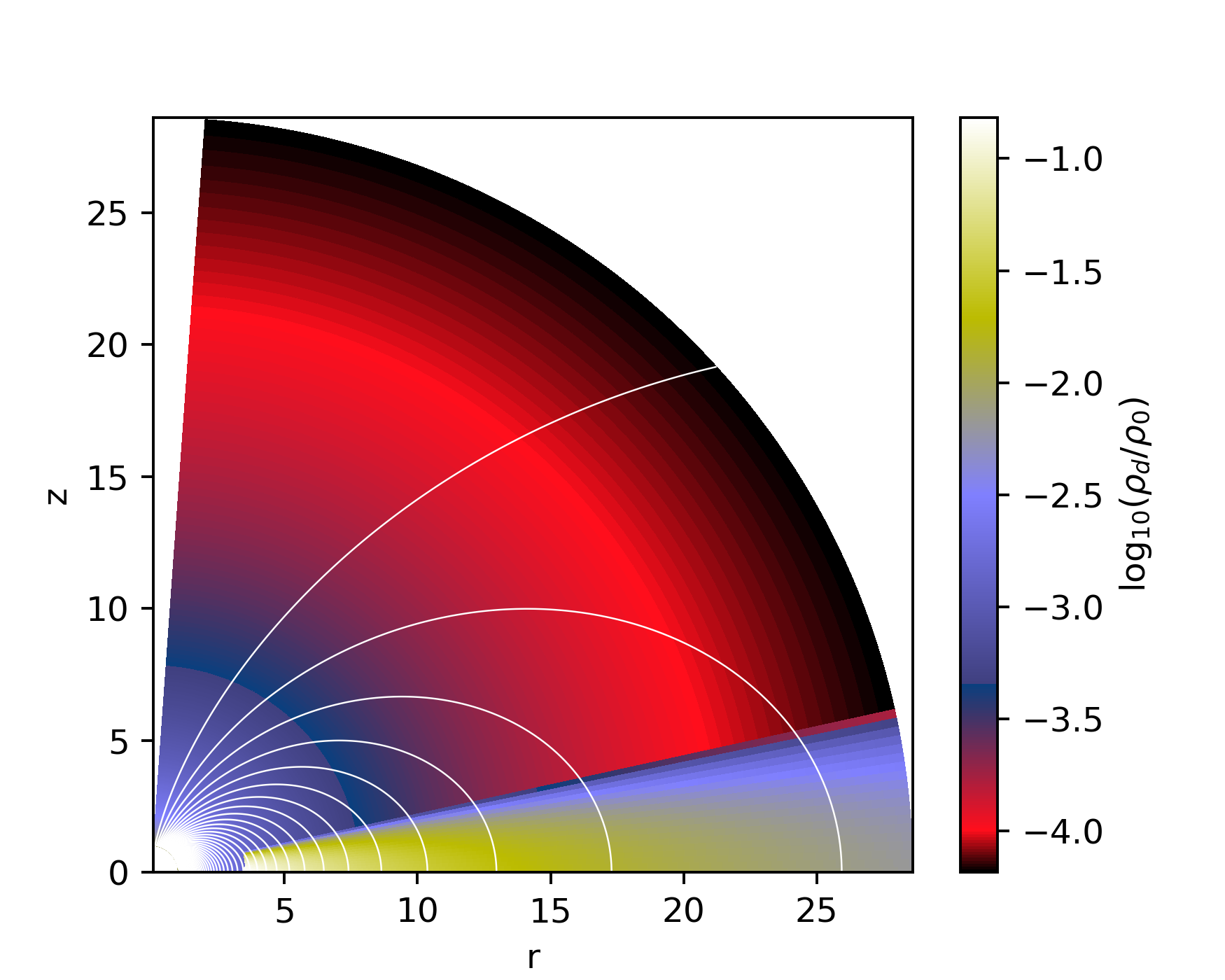}
    \caption{Initial density in the model DTHICK$\_13$ in logarithmic scale in units of $\rho_{0}$.   Sample field lines of the initially dipolar magnetosphere are also shown (white curves).}
    \label{fig:init}
\end{figure}


\subsection{Disc atmosphere}

In a similar form to \citet{Zanni2009}, we included a non-rotating politropic  hydrostatic atmosphere with a density 
\begin{equation}
    \rho_{\mathrm{atm}} (R)=\rho_{\mathrm{atm}}^0\left(\frac{R_\star}{R}\right)^{\frac{1}{\gamma-1}}
\end{equation}
and pressure
\begin{equation}
    P_{\mathrm{atm}}(R)=\rho_{\mathrm{atm}}^0\frac{\gamma-1}{\gamma}\frac{GM_\star}{R_\star}\left(\frac{R_\star}{R}\right)^{\frac{\gamma}{\gamma-1}}
\end{equation}
with $\gamma =5/3$.
The density contrast between the disc and the atmosphere is $\rho_\mathrm{atm}^0/\rho_{d0}=0.01$, which is kept fixed in all our models.

\subsection{Magnetic field configuration}
We assume that the star is located at the origin of the coordinate system.
The stellar magnetosphere is modeled initially as a purely dipolar field aligned with the rotation axis of the star-disc system. The magnetic field is defined by the flux function $\Psi_*$
\begin{equation}
    \Psi_*(R,\theta)=B_*R_*^3\frac{\sin^2{\theta}}{R}
\end{equation}
where $B_*$ is the magnetic field at $R_{*}$ and $z=0$. The radial and polar field components are therefore given respectively by

\begin{equation}
    B_R = \frac{1}{R^2\sin{\theta}}\frac{\partial\Psi_*}{\partial\theta}
\end{equation}
and
\begin{equation}
    B_\theta = -\frac{1}{R\sin{\theta}}\frac{\partial\Psi_*}{\partial R}.
\end{equation}
The relation between flux function and potencial vector is given by:
\begin{equation}
    \Psi_*=RA_\phi\sin{\theta}
\end{equation}
and then the components of potential vector in spherical coordinates are given by:
\begin{equation}
    A_\phi(R,\theta)=\frac{B_*R_*^3\sin{\theta}}{R^2},
\end{equation}
 $A_R=0$ and $A_{\theta}=0$. In Fig. \ref{fig:init} we show the initial distribution of the gas and corona densities, as well as the initial magnetic field configuration. 

\section{Numerical Implementation}
\label{sec:numerical}
\subsection{Code}

In order to investigate the disc-star interaction, we use PLUTO code (\cite{Mignone2009}, \cite{Mignone2007}) to perform adiabatic global MHD simulations of accretion discs to numerically solve the MHD equations given above. Our computations run on spherical geometry $(R,\theta,\phi)$ using the HLL Riemann solver and the 2nd-order Runge Kutta
integration in time to advance conserved variables. 

\subsection{Simulation parameters and units}
The parameters of the simulations were selected to match a set of observed values for the FS~CMa stars
\citep{Korcakova22}. 
More specifically, we fix the mass, radius, and the magnetic field of the star to be $M_*=6M_\odot$,
$R_*=3R_\odot$, and $B_*=6.2\times10^{3}G$.
Since previous studies have found that the local density around FS~CMa type stars is very low,  $\rho \simeq 1\times10^{-11} \mathrm{g\, cm^{-3}}$ \citep{Kricek}, or probably
below $\rho \simeq 1\times10^{-12} \mathrm{g\, cm^{-3}}$ \citep{Korcakova22},
we have carried out simulations varying the initial disc density between $[10^{-13}\mathrm{g\,cm^{-3}},10^{-11}\mathrm{g \, cm^{-3}}]$.
In addition, we have considered two different aspect ratios $h =0.05$ and $h=0.1$. The different models considered are presented in Table \ref{tab:Tabla1}.

We use the following dimensionless variables:

\begin{align*}
    R^{\prime}&=R/R_0 & \rho'& = \rho/\rho_{0}  & v'& =v/v_{K0} \\
    t'&=t/(r_0/v_{K0}) & p'&=p/(\rho_{0}v^2_{K0}) &          
B'&=B/(4\pi\rho_{0}v^2_{K0})^{1/2}
\end{align*} 
with
$R_0=R_*=2.087\times 10^{11}\mathrm{cm}$, $v_{K0}=\sqrt{GM_*/R_0}\approx 6.19 \times 10^7 \mathrm{cm\ s^{-1}}$, which is
the Keplerian velocity at $R_0$. For the reference density we take $\rho_{0} = 1\times10^{-13} \mathrm{g\, cm^{-3}}$.
 
In these units, the reference angular rotation rate is $\omega_0=v_0/R_0 =2.96\times 10^{-4}\mathrm{s^{-1}}$, the corresponding timescale is
$t_0= R_0/v_0 \approx 3.37\times 10^3\mathrm{s}$ and the rotation period at $R =R_0$, $T_0=2\pi t_0 \approx 0.245$ days.
For simplicity, we will omit the primes from now on.

\begin{table}
    \centering
    \begin{tabular}{p{1.5cm}| p{0.8cm}| p{2.0cm}| p{1.0cm}|}
        \hline
        Model & $h$ & $\rho_{d0}[\mathrm{g}\,\mathrm{cm}^{-3}]$& $B_\star[G]$ \\ \hline
        DTHIN$\_13$& $0.05$&$1\times 10^{-13}$& $6200$ \\
        DTHICK$\_13$& $0.1$&$1\times 10^{-13}$ & $6200$ \\
        DTHIN$\_12$& $0.05$&$1\times 10^{-12}$ & $6200$\\
        DTHICK$\_12$& $0.1$&$1\times 10^{-12}$& $6200$ \\
        DTHIN$\_11$& $0.05$&$1\times 10^{-11}$ & $6200$ \\
        DTHICK$\_11$& $0.1$&$1\times 10^{-11}$& $6200$ \\
        \hline
    \end{tabular}
    \caption{Model parameters of our numerical models.}
    \label{tab:Tabla1}
\end{table}

\subsection{Mesh domain and boundary conditions}

Our numerical models run on a CPU cluster, therefore, the computational cost of simulating a full disc would have prohibitive computational cost, especially for parameter space explorations presented here (see Table \ref{tab:Tabla1}). For this
reason we restrict our computational domain in the three
coordinates. In the $R$-direction we cover the interval
$[R_\mathrm{min},R_\mathrm{max}]=[1.0R_0,26.8R_0]$, in the $\theta$-direction\footnote{Since RING AVERAGE flag that helps to remove the CFL restriction near the pole/axis, which is based on the ring averge technique \citep{Zhang19} not work with constrained transport MHD, then we use $\theta_\mathrm{{min}} \neq 0$.} we use
$[\theta_\mathrm{min},\theta_\mathrm{max}]=[\frac{\pi}{44},\frac{\pi}{2}]$. Finally, since we are considering axisymmetric models, we do not include an explicit azimuthal extent. Therefore, cell numbers in each direction are
$(N_R,N_{\theta},N_{\phi})=(846,400,1)$.

The boundary conditions implemented in our models in the radial direction are of the zero gradient type for the scalar quantities $(\rho,e,P)$. While for vector quantities such as velocity and magnetic field, we use boundary conditions similar to those given in \citet{Romanova_etal2002}. For instance, for the magnetic field the boundary condition at $R_{\rm min}$ is $\partial (RB_\phi)/\partial R= 0$, and for the $B_R$ and
$B_{\theta}$ components are derived from the magnetic flux function $\Psi_\star(R, \theta)$. On the other hand, at the outer boundary $R = R_{\rm max}$, we take free boundary conditions for all quantities. Lastly, in the $\theta$-direction we use outflow boundary conditions at $\theta=\theta_\mathrm{min}$, and since we simulate only one hemisphere of the disc we use reflecting boundary conditions at the midplane (at $\theta=\theta_\mathrm{max}$).

\section{Results}

\label{sec:results}

\begin{figure}
\centering
\includegraphics[scale=0.42]{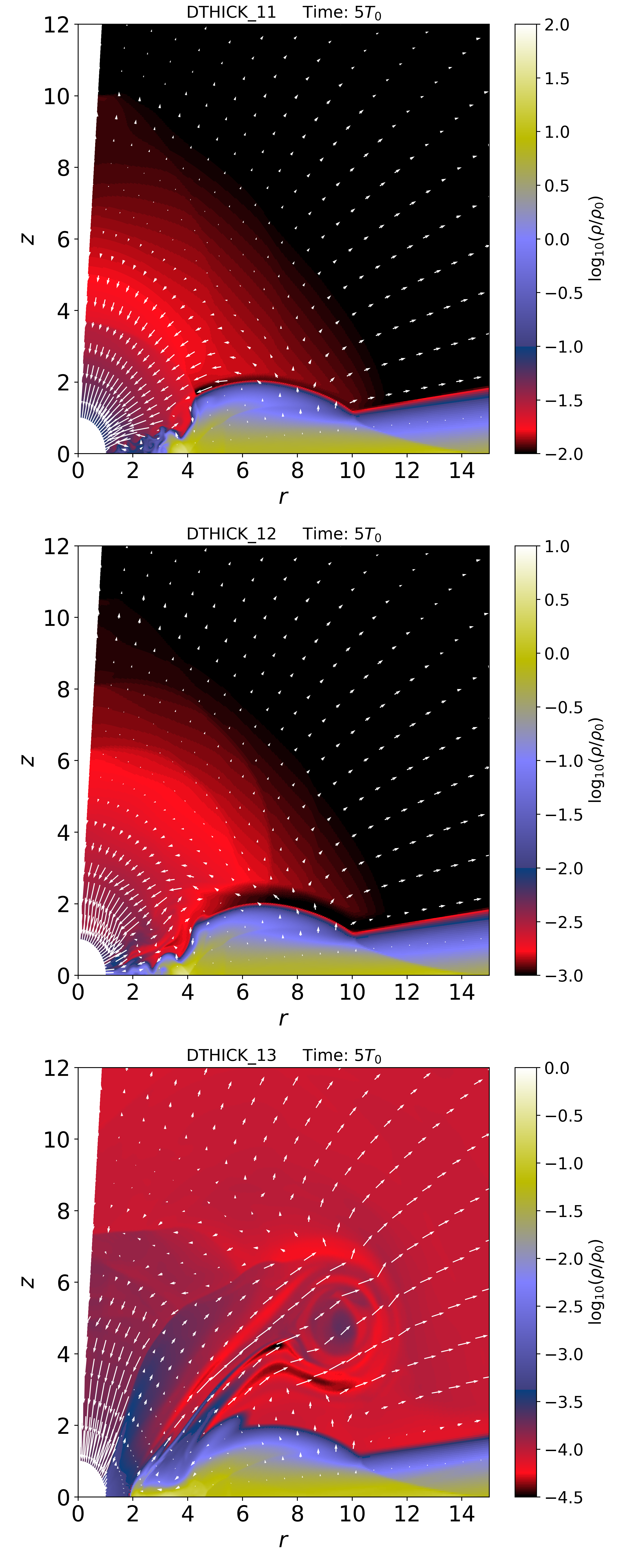} \caption{Disc and corona density for three different models (see Table \ref{tab:Tabla1}) at $t=5T_0$. White arrows depict velocity vectors in the $R-z$ plane. Note, the color scale is different
from panel to panel in order to see the structures in the corona region, since it is where the "optical-jet" or "hot-plasmoid" can be form.} 
\label{fig:density}
\end{figure}

Figure \ref{fig:density} shows the density of the circumstellar disc (which is not corotating with the central star) and the corona at a time of $t=5T_0$. The velocity vectors in the $R-z$ plane are also shown. In the DTHICK$\_$11 model, where the disc has the highest density, it can be seen that the coronal gas is falling towards the central star. A density bump is formed in the disc between $R=3.5$ and $R=10$. In model DTHICK$\_$12 an outflow in the disc layer in a radial interval from $R=2.5$ to $R=5$ starts to counteract the "free fall" of the coronal gas towards the central star. As in model DTHICK$\_$11, a density bump is observed
in the disc. In model DTHICK$\_$13, the accretion onto the star occurs mainly at a higher altitude near the polar axis within a relatively narrow region. In the bottom subplot Fig. \ref{fig:density} it can be seen the formation of a jet and a turbulent region, where the gas flow is outward. Several filaments of gas can also be seen emerging from the inner edge of the disc following the path of the outflow.

\begin{figure*}
	\includegraphics[scale=0.42]{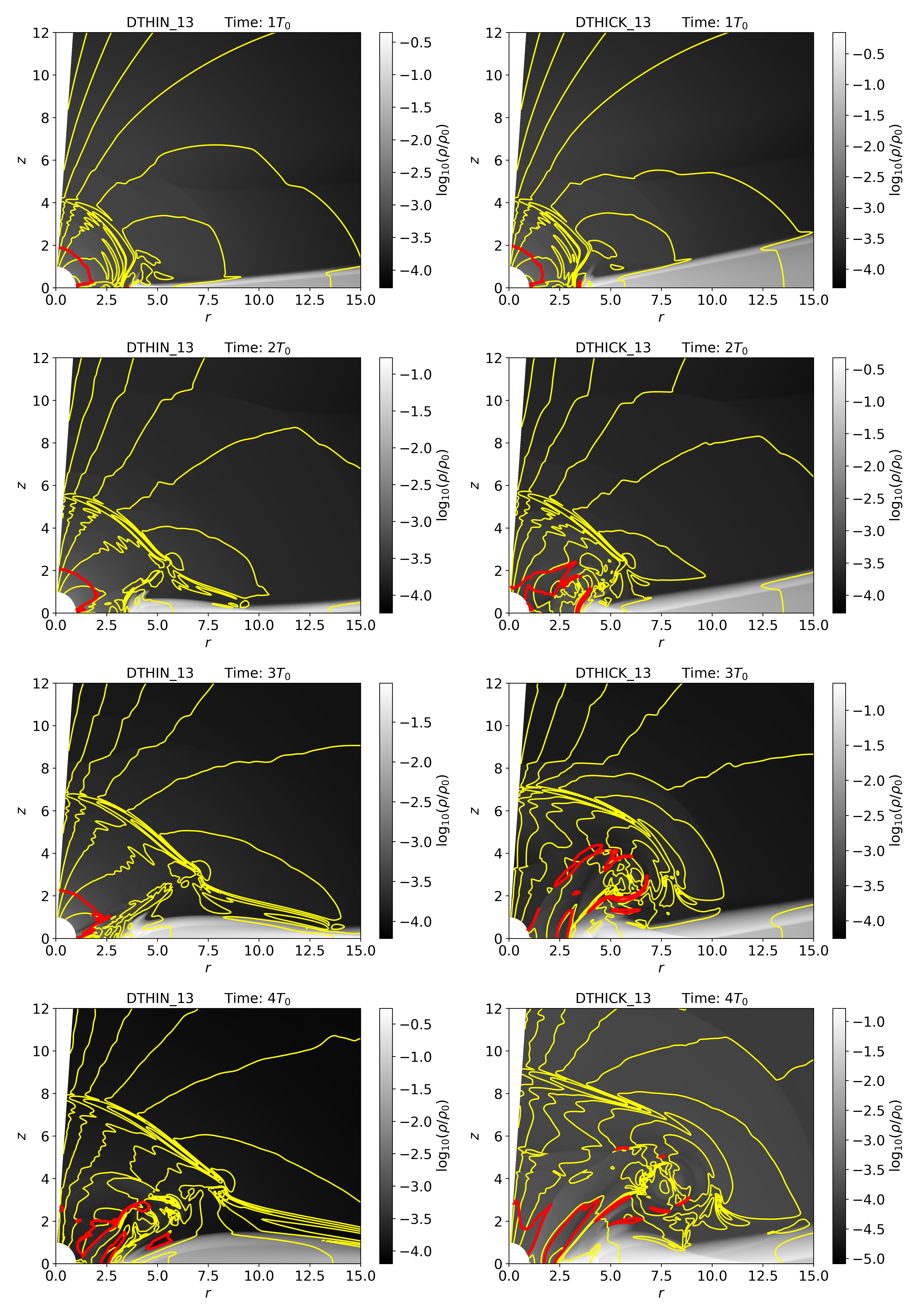}
    \caption{Temporal evolution of the gas density (gray-scale background) for the model DTHIN$\_13$ (left panels) and for model DTHICK$\_13$ (right panels). The time are shown at top of each panel which are measured in periods of Keplerian rotation. The yellow lines represent the poloidal field lines, the contours are exponentially spaced between $10^{-7}$ to $10^{-2}$. The red line corresponds to $\beta = 1$.}
    \label{fig:evolution_rho}
\end{figure*}

In the DTHICK$\_$11 and DTHICK$\_$12 models, the gas that flows to the
the central star persists in the equatorial region of the disc. However, we do not observe a "funnel effect" pattern like the one observed when the accretion is of the stationary type. \citep[see][]{,Bessolaz07,Romanova_etal2002,Zanni2009}. 
Nevertheless, we do not rule out that in these models an "intermittent funnel effect" may occur very close to the midplane of the disc. On the contrary, for the DTHICK$\_13$ model it is clearly seen that the fall of the gas towards the star in the region of the midplane of the disc is suddenly stopped by the outflow, thus delimiting the inner edge of the disc.

\begin{figure}
    \includegraphics[width=\columnwidth]{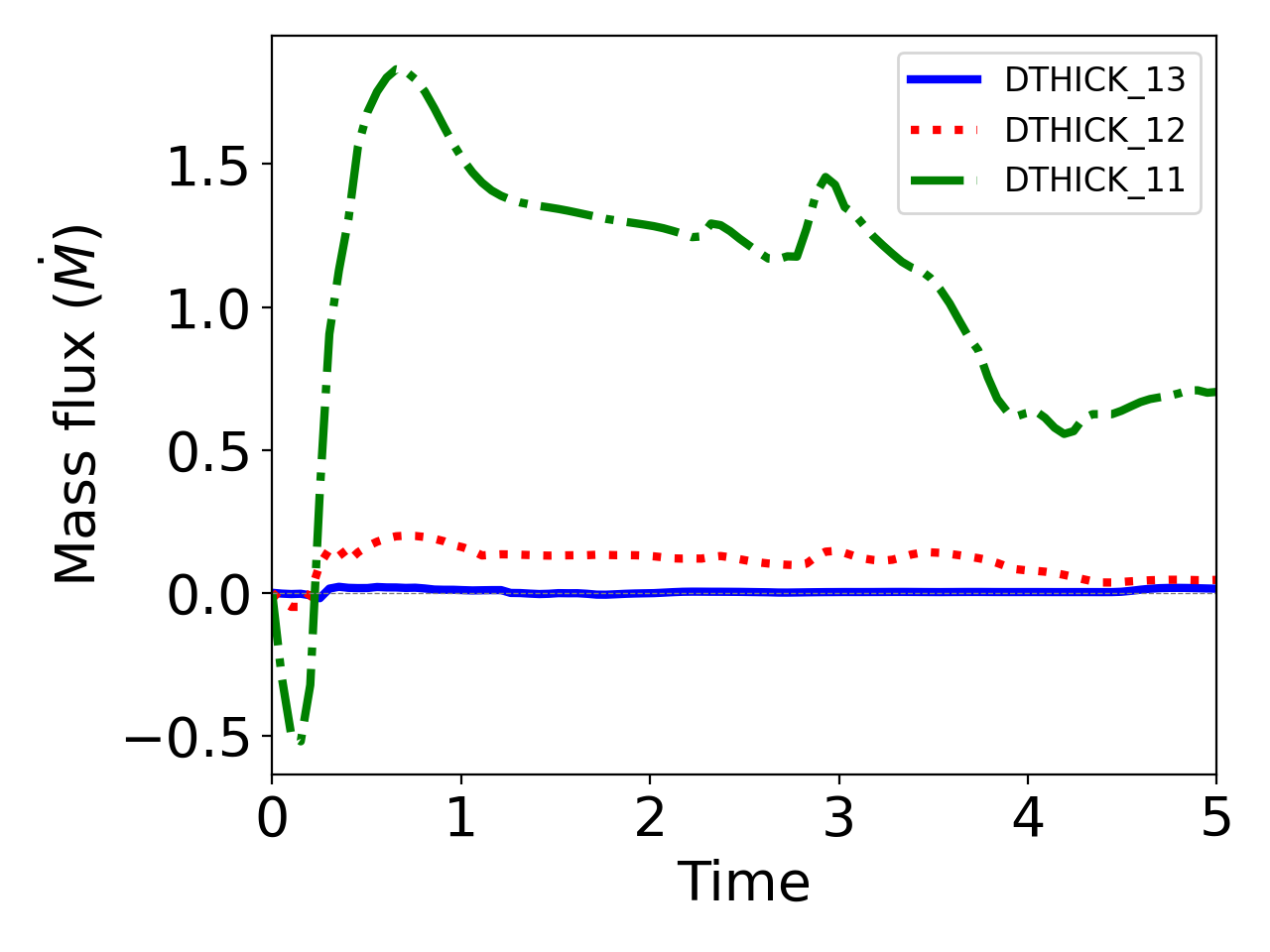}
    \caption{Time dependence mass flow rate $\dot{M}$, for the DTHICK models (see labels in the figure and Table \ref{tab:Tabla1}), calculated at $R=2R_0$.}
    \label{fig:mass_flux}
\end{figure}

\begin{figure}
    \centering
    \includegraphics[scale=0.42]{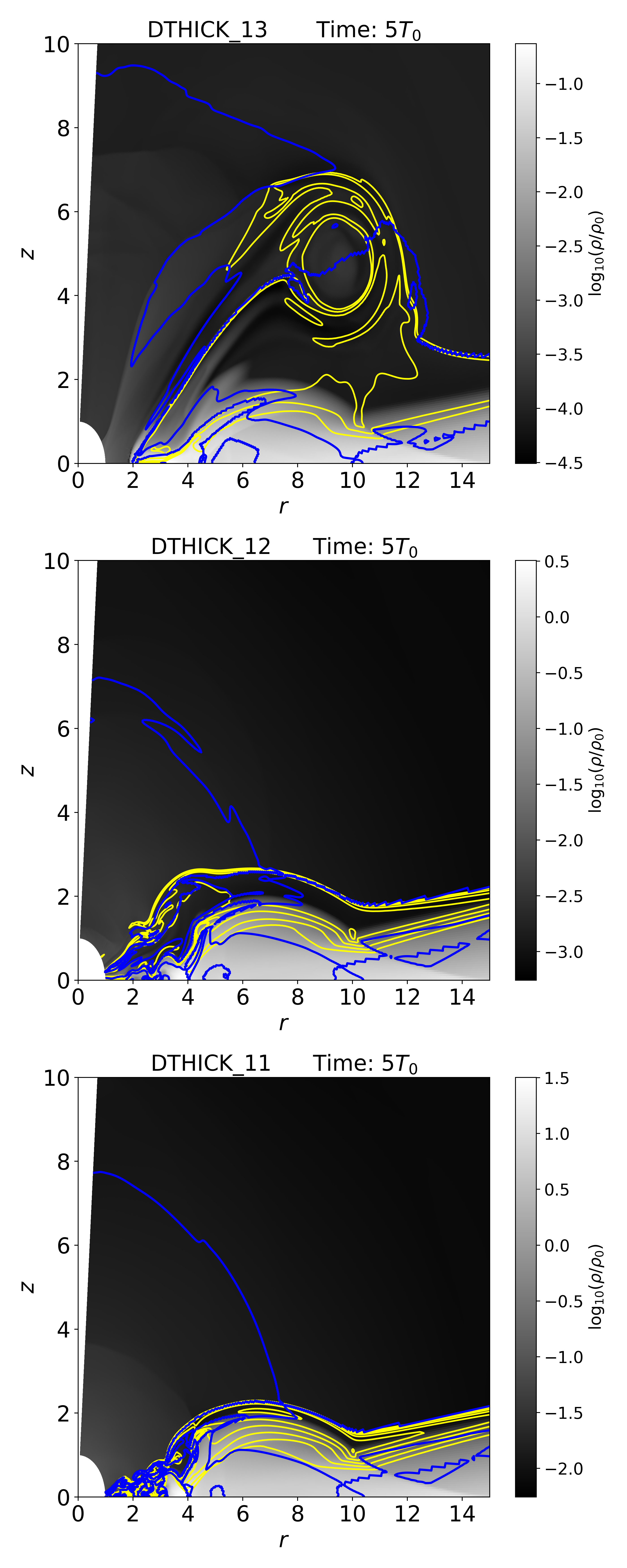}
    \caption{Poloidal electric circuits flowing in the star-disc-wind system. The contours correspond to values of $rB_{\phi}=Const$. A logarithmic density map (in units of $\rho_{d0}$) at \textbf{$T=5T_0$} is shown in the background, the top subfigure corresponds to the DTHICK$1\_13$ model, the middle subfigure  corresponds to the DTHICK$1\_12$ model and the bottom subfigure corresponds to the model DTHICK$1\_11$. The blue line shows the Álfven surface. }
    \label{alfven_contour}
\end{figure}

Fig. \ref{fig:evolution_rho} 
shows the temporal evolution of the gas disc density for the DTHIN$\_13$ and DTHICK$\_13$ models, yellow contours show poloidal magnetic field lines, while the red line shows the plasma parameter $\beta=(p+\rho \mathbf{v}^2)/(\mathbf{B}^2/8\pi) =1$ \citep{Romanova_etal2002}. The $\beta$-parameter is very useful to determine in which regions of the disc and corona the dynamics is dominated by magnetic stress. For instance, in disc-magnetized star interactions due to the differential rotation of the footpoints of magnetic field loops over the radial extension of the disc, produce the twisting of the magnetic field in the corona which can lead to open magnetic loops in some regions. To delimit those regions where this effect occurs, it is enough to know if the magnetic pressure is greater than the matter pressure, which happens when $\beta\leq1$ \citep{Romanova_etal1998}. In Fig. \ref{fig:evolution_rho} it can be seen that the magnetically dominated regions in our models are found at the inner edge of the disc and in the corona where the outflows take place.

On the other hand, we observe that the accretion disc is getting closer to the center as the time evolution progresses, starting at $T=1T_0$, around $R=3.6$ for both models, DTHICK$\_13$ and DTHIN$\_13$ (see first row of the figure \ref{fig:evolution_rho}),  whereas at $T=4T_0$ we found that the accretion disc stops at $R=2.3$  in DTHICK$\_13$ model, while the accretion disc in the DTHIN$\_13$ model stops at $R=2.6$. In other words, the accretion disc stopped at minor radii when the aspect ratio corresponds to $h=0.1$ (see last column of the figure \ref{fig:evolution_rho}). 

In addition, from Fig. \ref{fig:evolution_rho} we can see the temporal evolution of the density bump formed on the inner edge of the disc and the formation of a jet, which is more intense in the thicker disc. In both models, it can be seen that there is no gas fall towards the central star in the midplane of the disc. Basically, the accretion of gas towards the star occurs in the region close to the poles of the star. This result is important, since it means that the jet that forms in the corona, which originates from the inner edge of the disc, can be maintained for a longer period of time, since the outflows govern the dynamics between the corona and the layer of the disc avoiding the free fall of the gas.

On the other hand, the mass flow rate crossing the whole domain is the integral of the mass flux inside and outside of the disc. It is given by

\begin{equation}
    \Dot{M}= -4\pi R^2\int_0^{\pi/2} \rho u_{Rd} \sin{\theta} d\theta.
\end{equation}

The Fig. \ref{fig:mass_flux} shows the mass flow of the disc at $R=2.0R_0$. We find that when the density of the disc is higher the mass flow rate is positive for the most part, but shows a negative slope after $t=T_0$ (DTHICK$\_$11 and DTHICK$\_$12 models). This behaviour was expected since in these cases there is no formation of strong outflows the gas flow may be intermittent. For the DTHICK$\_$13 model we find that the rate of mass flux towards the central star can be negative or zero. This means that there is no inflow of gas towards the central star in the midplane of the disc and, therefore, this confirm that the accretion disc stop at a radius greater that $2R_0$.

\section{Discussion}
\label{sec:discussion}

The scenario studied in this first paper, where we consider that the accretion towards a central star of the FS~CMa type is non-stationary, is well justified since this accretion phase can mainly be triggered from a binary star post-merger. It should also be noted that there are different physical phenomena that can give rise to a non-stationary accretion, such as, thermal instability \citep{Lightman74}, global magnetic instability of the disc (\cite{Lovelase94}; \cite{Lovelase97}).

Due to the dipolar configuration of the magnetic field, the gas in the corona and in the inner region of the disc must first be lifted against gravity force \citep{Bessolaz07}. However,
since the rotation speed of the star is neglected (which is well justified for this type of stellar objects), and since we include a very strong magnetic field in our models, the accretion may be depend on different factors. For instance, the strong differential rotation in the inner region of the disc (which is an upper limit with respect to stationary accretion models), can produce azimuthal magnetic breaking. Then, the magnetic breaking "turns off" the centrifugal force, thus leaving the dynamics to be governed by the force of gravity. But on the other hand, due to the strong magnetic
field (in addition to a plasma pressure gradient in the inner region of the disc) it counteracts the force of gravity in greater proportion towards the surface of the disc ($\theta\rightarrow \pi/2$) ,
so the force of gravity mainly governs the accretion to a greater height of the midplane of the disc (toward the poles, see Fig. \ref{fig:density}).
On the other hand, it is important to emphasize that the dynamic evolution of the corona and the accretion disc, as we saw in the previous section, is governed mainly by the free fall gas from the corona and the density of the disc. When the disc has a lower density, there may be formation of different substructures in the corona-region (see bottom subplot Fig. \ref{fig:density}). These substructures can be identified as possible "optical-jets" or "hot-plasmoid" regions which have been previously observed in numerical simulations of accretion discs around young stars with a lower magnetic field intensity \citep[see][]{HSM1996}.
Furthermore, in Figure 3 we show the poloidal magnetic field lines for both
DTHIN and DTHICK models. It can be seen that due to the
non-corotating velocity between the star and the disc, the magnetic field lines twisted in a very short time. In fact, we can observe that since the time $1T_0$ which corresponds to an orbital period, the magnetic field lines have expanded or inflated. This configuration, in which the field of the open disc at the inner edge of the disc is in the opposite direction to the open stellar field, it favors the magnetic reconnection that can be observed both in the DTHIN model and in the DTHICK model as the small yellow circles in the plots. However, we find that magnetic
reconnection occurs at the inner edge of the disc and at the corona
region where the hot plasmoid forms.

One way to implicitly infer the effect of the magnetic field on the gas dynamics in the disc and corona region is through poloidal electric currents \citep{ZFRB2007,Zanni2009,MFZ2010}. With this goal, we show in Fig. \ref{alfven_contour} the gas density in the disc and corona regions for the models DTHICK$\_$11, DTHICK$\_$12 and DTHICK$\_$13, respectively at $t=5T_0$. We overlay  the poloidal currents, which are flowing along the isosurfaces defined by $rB_{\phi}=C$, with $C$ a constant. The poloidal currents circulate clockwise. Let us first analyze the poloidal currents in the case of the DTHICK$\_$13 model since in this model the disc density is the lowest considered in this study. From the top panel of Fig. \ref{alfven_contour} it can be seen that the current flows out of the inner boundary (from the star), and pass through the disc layer to the corona region forming closed current lines at different altitudes away from the disc midplane and within of the optical-jet. This behavior in the current lines is mainly governed by the azimuthal component of the magnetic torque $F_\phi=J_zB_r-J_rB_z$ which brakes the disc. Specifically, when $F_\phi>0$ at the disc surface (which means that $J_r$ decreases vertically), this torque provides a magnetic acceleration \citep{F1997} leading inevitably to an ejection. In the other two models (DTHICK$\_$12 and DTHICK$\_$11) the current lines describe a different behavior. In this case, the current lines emerge from the star and flow over the disc layer, said behavior can be handled by the differential rotation between the star and the disc. On the other hand, the non-spherical shape of the Alfven surface, shown by the blue countur in the figure \ref{alfven_contour}, can be due to magnetic rotational effects. Furthmore, we can see that the heigth over the z-axis increases that can explained  becouse the strength of the poloidal field there increases due to collimation \citep{Washimi93,Sean}.

Additionally, we have analyzed the backflow in the disc midplane for the DTHICK$\_$11, DTHICK$\_$12, and DTHICK$\_$13 models. The figure (\ref{fig:comp_vr}) shows the radial velocity component $u_{Rd}$ (calculated at $R=2R_0$) as a function of the time. We can see that in the cases when the density of the disc is higher (DTHICK$\_$11 and DTHICK$\_$12 models) the radial component of the gas velocity is negative in a short time, while for the case of the DTHICK$\_$13 model it is positive, this implies that the outflow occurs faster in the case of a low-density disc.

\subsection{Relation between simulations and observations}

The results found in our 2.5D-MHD simulations can explain several observed phenomena which we detail below.

\begin{figure}
    \centering
    \includegraphics[width=\columnwidth]{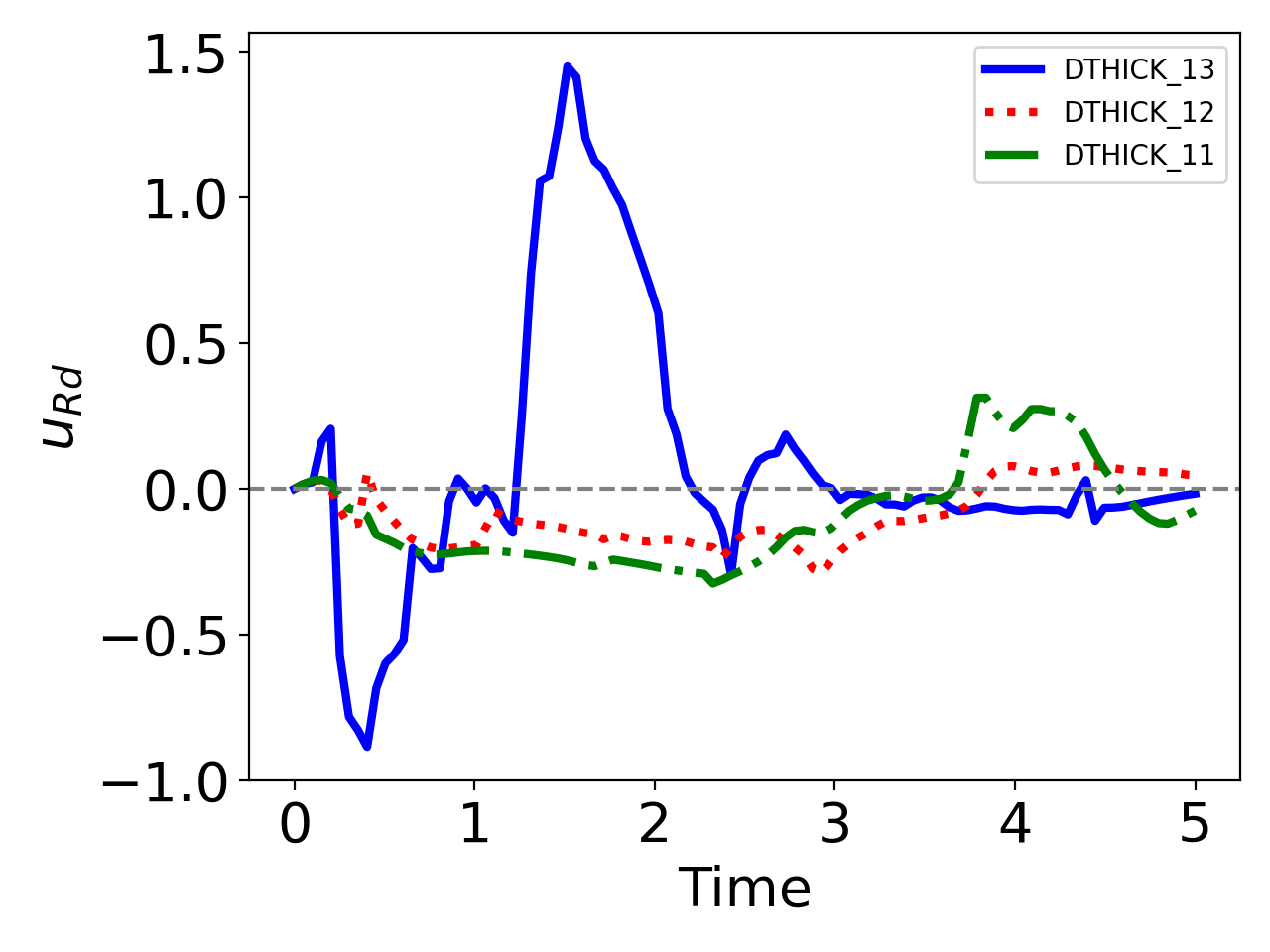}
    \caption{Radial velocity component as a function of time, for DTHICK models (see labels in the figure and Table \ref{tab:Tabla1}), calculated at radial position $R=2R_0$ over the midplane of acretion disc.}
    \label{fig:comp_vr}
\end{figure}

\begin{itemize} [leftmargin=*]
    \item \textit{Geometrically thick disc}. We performed two types of models see (Table \ref{tab:Tabla1}) with different height scales. We found that the thick model in the low-density regime match observations better. 
    \item \textit{Structured disc}.  In all our models resembles different structures as pressure bumps and cavities (Fig.~\ref{fig:evolution_rho}).
    \item \textit{Material infall}.
      The currents toward the stars created by the magnetic field in the low-density plasma may be responsible for the material infall. Such event 
      was detected in MWC~342 through the inverse P-Cygni profile \citep{Kucerova13}. 
    \item \textit{Discrete absorption components of resonance lines $\equiv$ material ejecta}.
      Resonance lines are formed in the corona and appear almost always in the emission. 
      As the transition probability of the resonance lines is very large, they allow the 
      detection of a relatively small gain of the material. The discrete absorption
      components were observed in HD~50138 \citep{Pogodin97}, FS~CMa \citep{deWinter97}, and 
      GG Car \citep{Marchiano12}. Such moving material fits to jets created in our 
      low-density models. 
    \item \textit{Clumps moving away}.
      The magnetic field configuration in the low-density plasma region favors the appearance of magnetocentrifugal winds (see Fig.~\ref{fig:density}), offers the explanation of
      the moving humps through the line profiles that are observed in almost every FS~CMa 
      star.
    \item \textit{Hot spot}. 
      The variable hot continuum source was found in IRAS~17449+2320 \citep{Korcakova22}
      based on the spectral fitting and line-intensity variations. Another hot source may be the corona. However, the corona is heated by enormous amount of small recconections rather than by a large one. The variability of the corona is connected with the disc, magnetic field, and stellar properties, therefore its variability on a scale affecting the stellar radiation should be on the longer time-scales.
    \item \textit{Corona}.
      Our simulations show a significant amount of the material out of the disc. Even if
      our models are without the inclusion of heating and cooling up to now, it is 
      reasonable to assume that this matter is very hot due to the magnetic reconnection.
      Indeed, several FS~CMa stars show weak Raman lines. There are probably not symbiotic 
      stars because of the lack of HeII lines. Therefore
      a~white dwarf is probably not responsible for the high energy photons and hot
      coronal region seems to be a reasonable explanation of the observed properties. 
    \item \textit{Moving decelerated layers}. The decrease/increase in the mass flux at the near boundary of the star (see Fig.~\ref{fig:mass_flux}), as well as the time-varying behavior of the radial component of velocity (see Fig. \ref{fig:comp_vr}) found in our numerical models can offer the explanation of the line-profile changes in MWC342 \citep{Kucerova13}.
    \item \textit{Mass-loss rate.} The velocity profile in the Fig.~\ref{fig:comp_vr} has an important consequence for the determination of the mass-loss rate. The mass-loss rate is possible to be found only based on the comparison of the observed and synthetic spectrum. It has been derived for three FS~CMa type objects: HD~87643 \citep{Pacheco82}, AS~78 \citep{Miroshnichenko_AS_78}, 
and IRAS 00470+6429 \citep{Carciofi10}.
All used models assumed a freely expanding medium. Our results confirmed the previous observations of 
\cite{Kucerova13} showing that the velocity law is more complex, far from 
the smooth one and far from the stable outflow. Especially 
dynamic is the lower density model (see Fig.~\ref{fig:comp_vr}) 
that fits the observations best. Under such conditions, 
the standard methods of the mass-loss rate determination give the highly 
overestimated result. Therefore, further calculations of synthetic spectra must take into account more complex velocity field.

    \item 
       \textit{Long-term variability}. We present in this paper only one snapshot of the evolution of the post-merger that is connected to the phase where the envelope is already transparent because it is possible to connect the simulations with the observations of IRAS 17449+2320. However, the description of the long-term behavior is more complex. The magnetic field slows down the star very effectively. As slower the stellar rotation as weaker is the flow on the star and it is more concentrated toward the equator. This effect will be described in more details in the following paper (Moranchel et al, in preparation).
\end{itemize}

\subsection{Comparison with previous models}

Since in astrophysics there are numerous systems composed of a magnetized star surrounded by an accretion disc, numerous studies have been developed \citep{HSM1996,MKS1997,Romanova_etal2002,Long_etal2005,Bessolaz07,Zanni2009}. In this work we have focused on the non-stationary accretion case to study the interaction between gas and a strong magnetic field in a system formed by magnetized star of the FS~CMa-type surrounded by an accretion disc and corona region. We have included in our model important constraints obtained from observations, such as the mass of the star, the strength of the magnetic field, and the rotation speed of the star \citep{Korcakova22}. These last two parameters can play an important role in the dynamics of the gas. It is important to note that stars of the FS~CMa's type exhibit a very low rotation speed (see Appendix \ref{sec:appendix}), so it is safe to consider a null stellar rotation in our models. On the other hand, these types of objects exhibit strong magnetic fields ($\sim6\times10^{3}$). With all of the above in mind, in this section we make a qualitative comparison with a previous study \cite{MKS1997} where different configurations of strong (weak) magnetic field and high (low) disc density in X-ray binary pulsars and classical T~Tauri stars were analyzed.
\citet{MKS1997} studied the stellar magnetosphere-accretion disc interaction considering three different initial configurations for the magnetic field: (1) a pure dipole stellar field, (2) a dipole stellar field excluded from the disc and (3) and a combination of a pure dipolar star field with a constant field in the axial direction. In each of these three models, they studied the effect of various physical paremeters on accretion onto  magnetized central star, as the intensity of the magnetic field, the density of the disc, the internal radius, the magnitude of the resistivity and the speed of stellar rotation, among others (see their Tables I, II and III). Particular attention should be paid to their models $I_f$ (rotating star) and $I_g$ (non-rotating star) of the case (1), since a configuration of the magnetic field similar to that of our models is considered. For these models, \cite{MKS1997} found that for a rapidly rotating star, the polar accretion is inhibited, and for a non-rotating star they found that the polar accretion can be maintained. Remarkably, in the latter case the accretion being highly dynamic. This result agrees with our results of the dynamic accretion observed in our low-density disc models (DTHIN models). Therefore, we think that the dynamics of the gas disc and the corona are mainly governed by the strong magnetic fields in FS~CMa's-type stars.

\section{Summary and Conclusions}
\label{sec:conclusions}
We analyzed the effect of the dipolar magnetic field from a non-rotating B type star on the inner rim of an accretion disc and on the corona of the disc, adopting a fixed initial value of $B_\star\simeq6\times10^3G$ extrapolated from the recent results obtained by observational study \citep{Korcakova22}. We consider two types of discs: 1) thin discs, with an aspect ratio of $h=0.05$ and 2) thick discs, with $h=0.1$. In both cases, we consider different values for the density of the disc that goes from $10^{-13}$ up to $10^{-11}$ $\mathrm{gcm}^{-3}$, which are values characteristic of this type of stars. Our main results can be summarized as follows:
\begin{itemize} [leftmargin=*]
    \item For more massive discs (higher 
    density, DTHICK$\_11$ and DTHICK$\_12$ models), we find that the effect of the magnetic field is only reflected at the inner edge of the disc producing a density bump of radial extent from $R=3.5R_0$ to $R=10R_0$. The evolution of the corona shows a drop free of the gas towards the central star.
    \item  For the cases of the lowest density of the disc studied here (DTHIN$\_$13 and DTHICK$\_13$ models) we find the formation of a jet in the corona region as well as a region called "hot plasmoid" \citep{HSM1996}, this jet originates from the inner edge of the disc.
\end{itemize}

From this last point we conclude that our findings may have direct implications for understanding the observational results of this type of stars, since the formation of this jet and the hot plasmoid region may generate different characteristics in the observed data.
The presented results may be important not only for the study of post-mergers among FS~CMa stars, 
but for all types of stars showing the B[e] phenomenon. The forbidden lines, which define this group,
are indicators of low-density material. In the presence of the magnetic field in such conditions, 
the polar jets, ``hot-plasmoid'', and currents toward the central star are created. 
Moreover, the disc structure is changed significantly, affecting the observed phenomena. 

\section*{Acknowledgements}

We are grateful to the referee for constructive and careful report, which significantly improved the quality of the manuscript. And also we thank Javier Sanchez-Salcedo for his very thoughtful comments and useful suggestions. The work of A.M.-B. and D.K. was supported by Charles University Research program (GA CR 17-00871S). The work of R.O.C. was supported by the Czech Science Foundation (grant 21-23067M). Computational resources
were available thanks to the Ministry of Education, Youth and Sports
of the Czech Republic through the e-INFRA CZ (ID:90140).
A.M-B. thank Pablo F. Velázquez (ICN-UNAM) for allowing us to use the cluster for our first tests and also thank Enrique Palacios-Boneta (c\'omputo-ICN) for maintaining the Linux servers where our first tests simulations were carried out.

\section*{Data Availability}

 The PLUTO code is available from \href{http://plutocode.ph.unito.it}{http://plutocode.ph.unito.it}. The input files for generating our 3D magnetohydrodynamical simulations will be shared on reasonable request to the corresponding author.



\bibliographystyle{mnras}
\bibliography{example} 




\appendix

\section{Rotation velocities of FS~CMa stars}
\label{sec:appendix}
Table \ref{vrottab} summarises the measured values of central stars of FS~CMa group. Since these central stars
are embedded in the large amount of the circumstellar matter, the measured velocities may be affected by the emission originates
in the surrounding medium, flows present in the line-forming region, inaccurate guess of the limb darkening, or the splitting
of spectral lines in the magnetic field. Due to all listed effects, the real rotation velocity may be smaller.

\onecolumn
\begin{table}
\begin{tabular}{l|l|l|l|l}
 \hline
 IRAS ID    &  other ID        & rot. velocity     & method & notes   \\ \hline

  00470+6429 &   & low rot. vel.   & line width & \cite{Miroshnichenko_TUBP_III} \\ \hline  
  03421+2935 & MWC 728  &  $v \sin i \sim 110$ \kms           &  \ion{He}{i}, \ion{Mg}{ii} lines   & \cite{MWC728} \\ \hline        
  03549+5602 & 	AS 78         & $v \sin i \sim 70$ \kms           & \ion{He}{i} lines   &   \cite{Miroshnichenko_AS_78}\\ \hline
  06158+1517 &	MWC 137$^{C}$ &  $v_{\text{eq.}}<10$ \kms           & evolutionary models &   \cite{Michaela21_MWC137}\\ \hline
  06259-1301 &	FS CMa        &  $v_{\text{eq.}}<70$ \kms          & line profiles & \cite{Israelian96}\\ \hline
  06491-0654 &	HD 50138      & 90 \kms                          &  & \cite{Lee16}\\ \hline
  07080+0605 &                  & $v \sin i= 65\pm 2$  \kms         & \ion{Fe}{ii}, \ion{Si}{ii} lines        &	 \cite{Arias18} \\ \hline
  07370-2438 &	AS 160        & $v \sin i \sim 200 $ \kms         & fit of \ion{He}{i} 5876 \AA &  \cite{Miroshnichenko_AS160_03} \\ \hline
  07418-2850 &	3 Pup         & $v \sin i= 73$ \kms               & & \cite{Hoffleit82} \\\cline{3-5}
             &                  & $v \sin i= 50\pm5$ \kms           & line width & \cite{Plets95} \\ \cline{3-5}
             &                  & $v \sin i= 35\pm5$ \kms           & Fourier transform &  \cite{Miroshnichenko20_3Pup} \\ \hline
  08128-5000 & 	Hen3-140      & $v \sin i= 70$ \kms              &  comparison with SYNSPEC &   \cite{Miroshnichenko_warm_dust_01}\\ \hline
  09350-5314 &	Hen3-298      & not rapid rotator                & \ion{He}{i} line width & \cite{Miroshnichenko05_298+303} \\ \hline
  09489-6044 &	HD 85567      & $v \sin i= 50$ \kms              &  comparison with SYNSPEC &   \cite{Miroshnichenko_warm_dust_01}\\ \cline{3-5}
             &                  & $v \sin i= 73$ \kms               & Fourier transform & \cite{Khokhlov17} \\ \hline
  16031-5255 &	CPD-52 9243  & $v=36\pm 4$ \kms                  & line profiles \& interferometry & \cite{Cidale12} \\ \hline
  17117-4016 &	HDE 327083   & $v \sin i \sim 200$ \kms          &   &  \cite{Miroshnichenko03_HDE_327083} \\ \hline
  17213-3841 &	Hen3-1398    & $v \sin i \sim 50$ \kms          & comparison with SYNSPEC  &\cite{Miroshnichenko_warm_dust_01} \\ \hline
  20090+3809 &	AS 386        & $v \sin i=25\pm3$ \kms           & Fourier transform & \cite{Khokhlov18}         \\\hline 
  22248+6058 &	V669 Cep     & $v \sin i \sim 50$ \kms            & line profiles  &\cite{Miroshnichenko02_V669Cep}         \\\hline
  22407+6008 &	MWC~657      & $v \sin i \sim 100$ \kms            & line profiles  &\cite{Miroshnichenko_AS_78}         \\\hline
 
\end{tabular}
\caption{Measured rotation velocities of the central stars of FS~CMa stars. ($C$ denotes the candidate objects.)} 
\label{vrottab}
\end{table}
\twocolumn

%


\bsp	
\label{lastpage}
\end{document}